\documentclass[10pt]{iopart}
\usepackage{graphicx}

\begin{document}
\jl{3}	

\title[Volume--controlled buckling of spherical elastic caps]{Volume--controlled buckling of thin elastic shells: Application to crusts formed on evaporating partially--wetted droplets}

\author{D. A. Head}
\address{Department of Applied Physics, The University of Tokyo, Hongo 7-3-1, Bunkyo-ku, Tokyo 113-8656, Japan }

\date{\today}

\begin{abstract}
Motivated by the buckling of glassy crusts formed on evaporating droplets of polymer and colloid solutions, we numerically model the deformation and buckling of spherical elastic caps controlled by varying the volume between the shell and the substrate. This volume constraint mimics the incompressibility of the unevaporated solvent. Discontinuous buckling is found to occur for sufficiently thin and/or large contact angle shells, and robustly takes the form of a single circular region near the boundary that `snaps' to an inverted shape, in contrast to externally pressurised shells. Scaling theory for shallow shells is shown to well approximate the critical buckling volume, the subsequent enlargement of the inverted region and the contact line force.
\end{abstract}




The properties of fluid interfaces are strongly modified by the presence of a permeable solid layer~\cite{Binks:2002}. This is deliberately exploited in solid--stabilised or `Pickering' emulsions, where added colliodal particles become pinned at the liquid--liquid interfaces, inhibiting dewetting and vastly reducing droplet coalescence~\cite{Xu:2005,Asekomhe:2005}. Solid interfacial layers may also form spontaneously, as in the observation of a glassy `crust' on the surface of evaporating polymer or colloid solutions in thin film~\cite{deGennes:2002,Strawhecker:2001}, suspended droplet~\cite{Tsapis:2005} and partially--wetted droplet~\cite{Pauchard:2003b,Gorand:2004,Pauchard:2003a,Kajiya:pre,Pauchard:2003c,Pauchard:2004} geometries. In all of the above examples, it has been observed that, as the fluid volume(s) change with time, either by evaporation or draining (and might also be expected to occur during Ostwald ripening of emulsions), the interfacial area contracts, compressing the solid layer which then wrinkles or buckles as an elastic sheet. Predicting the onset and nature of this buckling is thus crucial to controlling system evolution and preventing the occurrence of undesired features in any given application.

The deformation and buckling of thin elastic shells is a classic problem; see {\em e.g.}~\cite{Seide,TovstikSmirnov,Shilkrut,LibaiSimmonds}. However, the systems mentioned above introduce a complication that has not, to the best of our knowledge, been properly treated, namely that the presence of the incompressible fluid  imposes a {\em volume constraint} on the space of allowed deformations (on time scales shorter than {\em e.g.} the evaporation time)~\cite{Pauchard:2003b}. It might be thought that controlling the volume $V$ would be identical to imposing some corresponding pressure~$P$, but this is {\em not necesarily} true with regards the {\em stability} of the shell: an S--shaped $P$--$V$ curve would reveal different limit points, and hence distinct buckling events~\cite{ThompsonHunt,Gilmore}, depending on which quantity is being controlled (similar to S--shaped flow curves in non--linear rheology, which may be stable under stress control but unstable under an imposed flow rate or {\em vice versa}; see {\em e.g.}~\cite{Head:2002}). 

\begin{figure}
\centerline{\includegraphics[width=10cm]{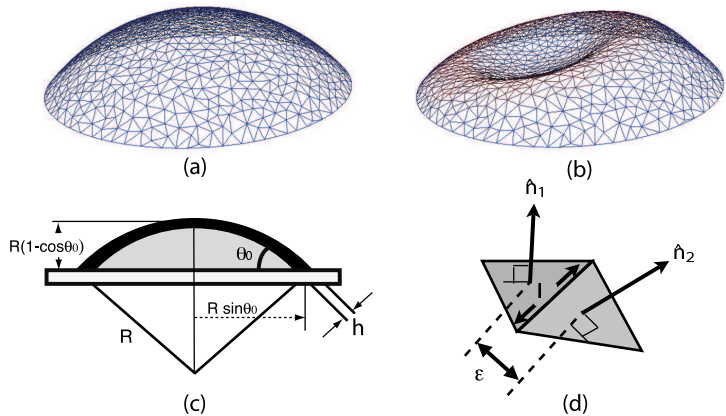}}
\caption{Examples of (a)~prebuckled and (b)~post--buckled shell
configurations ($\theta_{0}=0.8$, $h/R=0.03$). (c)~Undeformed geometry, a spherical cap of thickness~$h$, contact angle~$\theta_{0}$ and radius of curvature~$R$. (d)~Schematic of the discrete energy calculation (see text for details).
\label{f:intro}}
\end{figure}

Below we describe numerical simulations of a thin elastic shell with the same geometry as the crust on a partially--wetted spherical droplet, with the shell deformation driven by incrementally reducing the volume between it and the substrate. This geometry was chosen for its industrial relevance: the drying of inks and paints~\cite{deGans:2004,Zhang:1983} and  `bottom--up' manufacturing of microscale electronic devices {\em via} microfluidic technology~\cite{Cuk:2000} all involve the evaporation of partially--wetted solutions. Our primary finding is of a single class of buckling event for thin and/or steep shells, in which a region of the shell near the fixed boundary `snaps' to an inverted configuration [see Fig.~\ref{f:intro}], in contrast to the range of axisymmetry preserving and breaking buckling modes already documented for the pressure--controlled case~\cite{Shilkrut}. For shallow shells, the critical buckling volume $V_{\rm c}$ extracted from the simulations, as well as the subsequent enlargement of the inverted region, is shown to be well estimated by predictions of a simple scaling theory. We first describe the employed simulation method.

%
%
\vskip\baselineskip

\noindent{\em Methodology:} The crust is modelled as a thin elastic shell described by a 2--dimensional mid--surface $\mathcal{S}$ parameterised by a thickness~$h$, such that the crust surfaces lie at a distance $\pm h/2$ normal to $\mathcal{S}$. As the droplet surface is spherical, we take $\mathcal{S}$ to be a section of sphere of radius $R$ that intersects the substrate at an angle $\theta_{0}$; see Fig.~\ref{f:intro}(c). Deviations from this unstressed state are described by a pair of rank--2 tensors; $u_{ij}$, the in--surface or {\em membrane} strains, and the change in curvature $\Delta\chi_{ij}$ (with corrections for the difference between centroidal and mid--surfaces~\cite{Seide}). A standard closure due to Kirchoff and Love then gives the following expression for the strain energy, partitioned into a membrane (or `stretch') contribution $U_{\rm stretch}$, and a flexural (or `bend') term $U_{\rm bend}$~\cite{TovstikSmirnov},

\begin{eqnarray}
U_{\rm stretch}&=&\frac{1}{2}K\int{\rm d}\mathcal{S}\left\{Tr(u_{\alpha\beta})^{2}-2(1-\nu)det(u_{\alpha\beta})\right\},
\nonumber\\
U_{\rm bend}&=&\frac{1}{2}D\int{\rm d}\mathcal{S}\left\{Tr(\Delta\chi_{\alpha\beta})^{2}-2(1-\nu)det(\Delta\chi_{\alpha\beta})\right\},
\nonumber\\
&&K= \frac{Eh}{1-\nu^2}
\,,\quad
D=\frac{Eh^{3}}{12(1-\nu^{2})}\,.
\label{e:ctm}
\end{eqnarray}

\noindent{}where $E$ and $\nu$ are respectively the Young's modulus and Poisson ratio of the shell material in bulk. Note that this is constitutively {\em linear}; all non--linearities are geometric in origin, arising from the finite displacements implicit in $u_{ij}$ and $\Delta\chi_{ij}$.

Discretising (\ref{e:ctm}) with a regular mesh runs the risk of introducing undesired coupling between long--range mesh ordering and any buckling modes that may arise. To remove this possibility, we employ a statistically isotropic and homogeneous random mesh, formed by constructing a set of nodal points added with uniform probability on the spherical surface, with the constraint of a minimum distance between any two nodes. This was then Delaunay triangulated using {\em qhull}~\cite{qhull}.
Energy contributions for local deformations are calculated in a simple finite element scheme, similar to that of recent full sphere~\cite{Tamura:2004} and wrinkling~\cite{Kramer:1997} studies. Mesh edges are treated as Hookean springs of natural length $l_{0}$, so that $\delta U_{\rm stretch}=\frac{1}{2}\alpha(l/l_{0}-1)^{2}$ with $l$ the current length, $\alpha=Eh\delta A/(1-\nu^{2})$ and $\delta A$ the area assigned to this edge. The curvature between two adjacent faces, with unit normals ${\bf{\hat n}_{1}}$ and ${\bf{\hat n}_{2}}$ and whose midpoints are separated by a distance~$\epsilon$ [see Fig.~\ref{f:intro}(d)], is $\chi\approx\phi/\epsilon$ with $\phi=\arccos({\bf{\hat n}_{1}}\cdot{\bf{\hat n}_{2}})\ll1$. The energy contribution is then $\delta U_{\rm bend}=\frac{1}{2}\beta(\chi-\chi_{0})^{2}$, where $\chi_{0}$ is the spontaneous curvature and $\beta=h^{2}\alpha/12$ to ensure convergence to the continuum limit suggested by~(\ref{e:ctm}).

A series of quasi--static configurations for a given mesh is generated by minimising the total strain energy for incrementally smaller volumes $V$ by the non--linear conjugate gradient method, with the single scalar constraint~$V$~\cite{Bonnans}. Clamped boundary conditions were employed, in which the position and angle of boundary elements are fixed, as suggested by experimental observation of real crusts~\cite{Gorand:2004}. Interactions with the substrate were introduced by including an energy cost $U_{\rm sub}$ when shell nodes penetrate the substrate, with $U_{\rm sub}$ made sufficiently stiff that the shell is essentially flat when in contact, as confirmed by direct visualisation.

%
%

%
%
\vskip\baselineskip

\noindent{\em Results:} As the volume $V$ is reduced from its undeformed value~$V_{0}$, the shell initially undergoes axisymmetric deformations, with higher--order azimuthal modes ({\em i.e.} dimple $\rightarrow$ mexican hat $\rightarrow$ crater $\rightarrow$ ...) arising for thinner and/or steeper shells, much as in the externally--pressurised case~\cite{Head:2006}. However, buckling events, identified here as discontinuous drops in the total strain energy for an arbitrarily small volume change, are qualitatively different, taking the form of a single, circular region near the boundary that becomes inverted, as seen in Fig.~\ref{f:intro} (the proximity to the boundary is likely because it masks off part of the high--energy rim; see below). This has been observed even for contact angles $\theta_0$ up to $2\frac{1}{2}$ radians, but differs from full spheres where multiple inverted regions are expected~\cite{Quilliet:2006} and experimentally observed~\cite{Tsapis:2005}. 

To quantify these observations, it is convenient to consider shallow shells \mbox{$\theta_{0}\ll1$}, for which the two dimensionless parameters that define the geometry of the unperturbed shell, $\theta_{0}$ and $h/R$, can be combined into the single scalar

\begin{equation}
\lambda
=
\left[12(1-\nu^2)\right]^{\frac{1}{4}}
\sqrt{\frac{R\theta_{0}^{2}}{h}}
\quad.
\label{e:lambda}
\end{equation}

\begin{figure}[htb]
\centerline{\includegraphics[width=10cm]{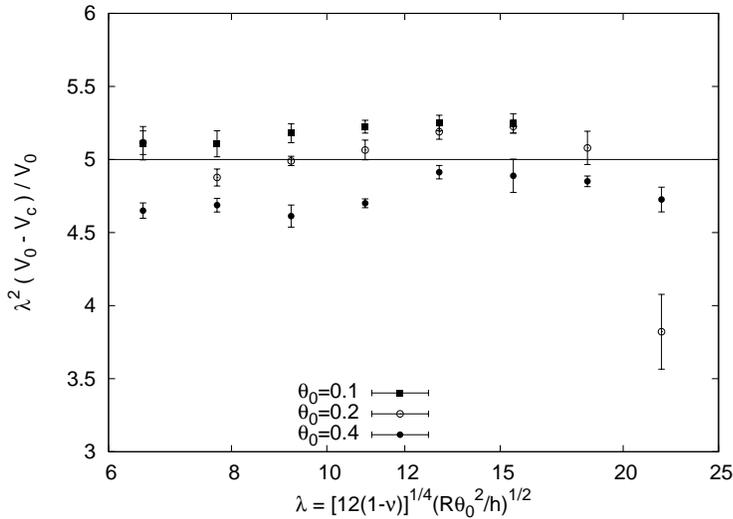}}
\caption{Critical volume scaled by $\lambda^{-2}$ {\em versus} $\lambda$ in the buckling regime $\lambda>\lambda_{\rm c}\approx5.25$ for the three different contact angles $\theta_{0}$ shown in the legend. The horizontal line corresponds to $\Delta V_{c}\sim\lambda^{-2}$ as predicted by scaling theory.\label{f:crit_V}}
\end{figure}

\noindent{}For externally pressurised shells, $\lambda$ is known to determine if a shell buckles, and the critical pressure at which it does~\cite{Shilkrut,Head:2006}; we show here that it plays a similar role under volume control. Buckling only occurs for $\lambda>\lambda_{\rm c}$ with $\lambda{\rm c}=5.25\pm0.15$, corresponding to thin and/or steep shells (note that this is significantly higher than the pressure controlled value $\lambda_{\rm c}^{\rm pressure}\approx3.3$~\cite{Shilkrut,Head:2006}). Furthermore, the dimensionless change in volume at which buckling occurs, $\Delta V_{\rm c}/V_{0}$, approximately scales as $\lambda^{-2}$ as in Fig.~\ref{f:crit_V}.

This relation can be understood by deriving approximate scaling relations between various characteristic quantities, in the style of~\cite{LandauLifshitz}. Let the characteristic radially--inward displacement be~$\zeta$ and assume this is the dominant mode. The undeformed height of the shell is $R(1-\cos\theta_{0})$ [see Fig.~\ref{f:intro}(c)], so $\Delta V/V_0\sim\zeta/R(1-\cos\theta_{0})\sim\zeta/R\theta_{0}^{2}$ for $\theta_{0}\ll1$. Employing the known result that buckling occurs when $\zeta\sim h$~\cite{Head:2006,LandauLifshitz}, we derive $\Delta V_{\rm c}/V_{0}\sim h/R\theta_{0}^{2}\sim\lambda^{-2}$, confirming the numerical findings. 

Extending this analysis to the strain energy provides insight into the physical meaning of the parameter~$\lambda$. Given the deformation mode described above, shell elements of area $\delta A$ will be compressed by a characteristic membrane strain $u_{ij}^{\rm char}\sim\zeta/R$. For bending, note that, in the pre-buckled state, the shell deforms smoothly over the entire surface, {\em i.e.} over a length $\propto R\theta_{0}$. Since the curvature is related to the second derivative of transverse displacements, $\Delta\chi_{ij}^{\rm char}\sim\zeta/(R\theta_{0})^{2}$. Inspection of (\ref{e:ctm}) then suggests characteristic energies $\delta U_{\rm stretch}\sim Eh(\zeta/R)^{2}\delta A$ and $\delta U_{\rm bend}\sim Eh^{3}[\zeta/(R\theta_{0})^{2}]^{2}\delta A$. The ratio is is $\delta U_{\rm stretch}/\delta U_{\rm bend}\sim R^{2}\theta_{0}^{4}/h^{2}\sim\lambda^{4}$, so $\lambda\gg1$ corresponds to a stretching dominated regime or {\em membrane state}, and conversely small $\lambda$ corresponds to a bending dominated response, somewhat comparable to flat plates.

%
%

\begin{figure}[htb]
\centerline{\includegraphics[width=10cm]{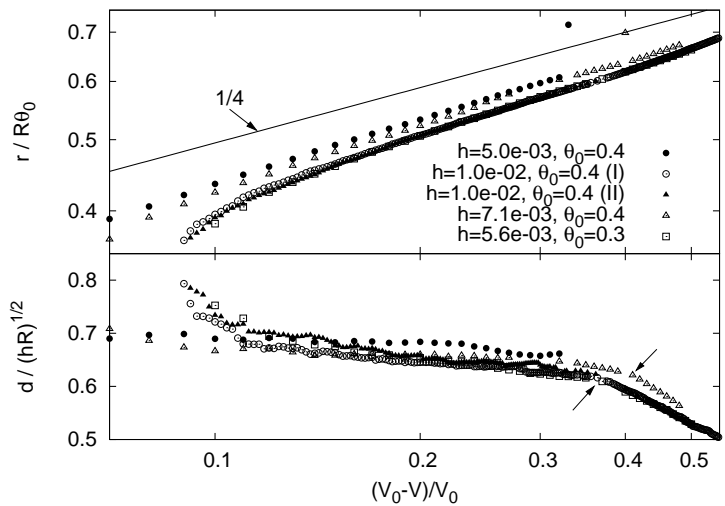}}
\caption{Evolution of the buckled region as a function of the relative reduction in volume $(V_{0}-V)/V_{0}=\Delta V/V_{0}$, found by fitting the strain energy density to the form $C e^{-(|{\bf x}-{\bf x_{0}}|-r)^{2}/2d^{2}}$ after stereographic projection to the horizontal ${\bf x}$--plane passing through the apex. The upper panel shows the radius $r$ of the inverted region, along with with the $1/4$ slope predicted by scaling theory. The lower panel gives the width $d$ of the high--stress 'rim' scaled by the prediction $d\sim\sqrt{hR}$ (the arrows indicate when the shell comes into contact with the base).\label{f:divet}}
\end{figure}

In the post--buckled state, the inverted region has roughly the same radius of curvature as the undeformed state but opposite sign; this gives almost zero stretch energy, and is thus energetically favourable for small~$h$. However, irrespective of~$h$, the rim between inverted and non--inverted regions is highly curved and the bending stresses can never be ignored here. With this insight, it is possible to extend the scaling analysis to the post--buckled state. It has been shown that the characteristic width $d$ of the rim that minimises the total energy scales as $d\sim\sqrt{hR}$~\cite{LandauLifshitz}. From purely geometric calculations, the volume of the inverted region of radius $r$ is $\Delta V/V_{0}\sim(r/R\theta_{0})^{4}$, thus scaling theory predicts $d\sim\sqrt{hR}$ and $r\sim R\theta_{0}(\Delta V/V_{0})^{1/4}$ in the post--buckled regime. This is confirmed by the numerical data in Fig.~\ref{f:divet}. Note that a parallel argument shows that the energy of $n$ identical inverted regions scales as $n^{1/4}$~\cite{Quilliet:2006}, thus the robustly observed $n=1$ case is indeed the energetically favoured one.


%
%

\begin{figure}[htb]
\centerline{\includegraphics[width=10cm]{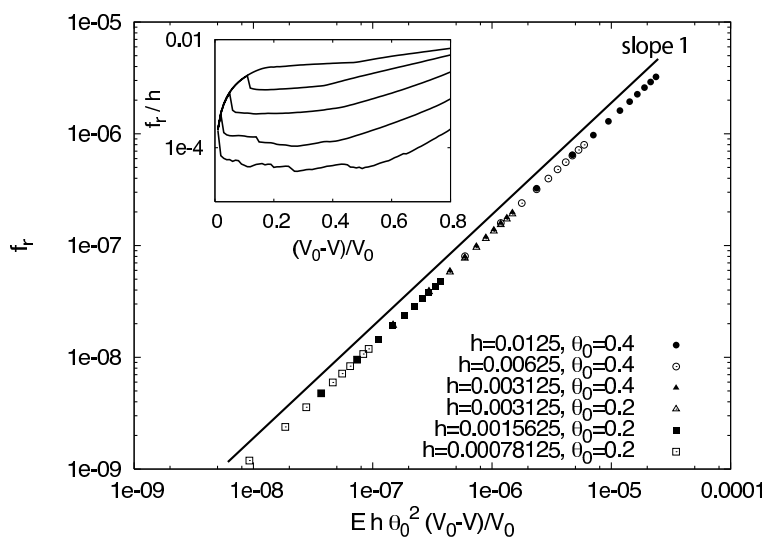}}
\caption{The radial component of the contact line force per unit length $f_{r}=f\cos(\theta_{0})\approx f$ versus $Eh\theta_{0}^{2}\Delta V/V_{0}$ for linear deformations, confirming the prediction of the scaling theory. {\em (Inset)}~Non--linear behaviour for $\theta_{0}=0.4$ and $h$ decreasing in factors of 2 from $h=0.025$ {\em (top)} to $h=0.0015625$ {\em (bottom)}. For all but the largest~$h$, buckling occurs at the discontinuity in the curve.
\label{f:residual}}
\end{figure}

In applications it is often important to know the forces at the contact line, as this can cause the crust to peel from the substrate, an undesirable effect in {\em e.g.} inks or paints. In the pre--buckled regime, the smoothly--varying membrane stresses scale with $Eu_{ij}^{\rm char}\sim E\zeta/R\sim E\theta_{0}^{2}\Delta V/V_{0}$, so the force per length of contact line is $f\sim Eu_{ij}^{\rm char}h\sim Eh\theta_{0}^{2}\Delta V/V_{0}$, which acts outwards since the shell is in compression. This is confirmed by the numerics; see Fig.~\ref{f:residual}. At the buckling volume $\Delta V_{\rm c}/V_{0}\sim h/R\theta_{0}^{2}$ (when $f_{\rm c}\sim Eh^{2}/R$), the contact force suddenly drops, as seen in the figure, where it remains of much lower magnitude until the shell becomes strongly deformed. Since this is now due to subdominant terms in the stress field, it is beyond the simple scaling theory presented here to predict its form.

For completeness, we have also simulated a sample of hinged caps where the contact angle is free to change. Although the buckling is of a similar form, the critical volume obeys $\Delta V_{\rm c}/V_{0}\sim\lambda^{-\alpha}$ with an exponent $\alpha\approx2.25$ rather than $2$ as for the clamped case; also, the post--buckled configurations interact with the substrate much sooner, nullifying the simple post--buckling analysis presented above.

%
%


%
%
\vskip \baselineskip

\noindent{\em Discussion:} The broad conclusion of this work is the inequivalence of pressure and volume control with regards the stability of thin elastic shells. For the spherical cap geometry considered here, volume control reveals just a single class of instability, contrasting with externally--pressurised caps~\cite{Shilkrut}. It is likely that a similar discrepancy will arise in other geometries, suggesting established buckling criteria will need to be rethought whenever incompressibility introduces a volume constraint to the space of allowed deformation modes.

Experiments on partially--wetted droplets have yet to show the buckling described here. This suggests that other mechanisms need to be incorporated, such as dynamics (solution and shell hydrodynamics and viscoelasticity), continued shell thickening during evaporation, and temperature gradients, amongst others. Indeed it is because of this the broad range of possibilities that a simple volume constraint was included in the simulations, rather than coupling to a uniform pressure field (the normal surface forces in the simulations are higher near the boundary); there seems little point demanding uniform pressure when the actual form will be far from trivial. It is hoped that a fuller model will soon be available to better understand this important phenomenon.

\vskip \baselineskip

\noindent{\em Acknowledgments:} The author would like to thank M. Doi and G. Fuller for useful discussions.
This work was funded by JSPS Fellowship P04727.


\end{document}